\documentclass[fleqn,usenatbib]{mnras}
\usepackage{newtxtext,newtxmath}
\usepackage[utf8]{inputenc}
\usepackage[T1]{fontenc}
\usepackage{ae,aecompl}
\usepackage{graphicx}	
\usepackage{amsmath}	
\usepackage{amssymb}	
\usepackage[normalem]{ulem}

\title[Models of RCB Stars]{Modeling R Coronae Borealis Stars: Effects of He-Burning Shell Temperature and Metallicity} 
\author[C. L. Crawford et al.]{
Courtney L. Crawford$^{1}$\thanks{Email: ccour14@lsu.edu},
Geoffrey C. Clayton$^{1}$,
Bradley Munson$^{1}$,
\newauthor 
Emmanouil Chatzopoulos$^{1}$,
 and Juhan Frank$^{1}$
\\
$^{1}$Dept. of Physics \& Astronomy, Louisiana State University, Baton Rouge, LA, 70803, USA\\
}
\date{Accepted 2020 August 15. Received 2020 August 14; in original form 2020 July 08}
\pubyear{2020}
\begin{document}
\label{firstpage}
\pagerange{\pageref{firstpage}--\pageref{lastpage}}
\maketitle

\begin{abstract}
The R Coronae Borealis (RCB) stars are extremely hydrogen-deficient carbon stars which produce large amounts of dust, causing sudden deep declines in brightness. They are believed to be formed primarily through white dwarf mergers. 
In this paper, we use {\it MESA} to investigate how post-merger objects with a range of initial He-burning shell temperatures from 2.1 - 5.4 $\times$10$^8$ K with solar and subsolar metallicities evolve into RCB stars. The most successful model of these has subsolar metallicity and an initial temperature near 3 $\times$10$^8$ K. 
We find a strong dependence on initial He-burning shell temperature for surface abundances of elements involved in the CNO cycle, as well as differences in effective temperature and radius of RCBs. 
 Elements involved in nucleosynthesis present around 1 dex diminished surface abundances in the 10\% solar metallicity models, with the exception of carbon and lithium which are discussed in detail.
Models with subsolar metallicities also exhibit longer lifetimes than their solar counterparts. 
Additionally, we find that convective mixing of the burned material occurs only in the first few years of post-merger evolution, after which the surface abundances are constant during and after the RCB phase, providing evidence for why these stars show a strong enhancement of partial He-burning products. 
\end{abstract}

\begin{keywords}
stars: abundances -- binaries: close -- stars: evolution -- white dwarfs
\end{keywords}

\section{Introduction}

R Coronae Borealis (RCB) stars are a rare type of cool supergiant star with severely diminished hydrogen and highly enriched carbon abundances \citep{1996PASP..108..225C,2012JAVSO..40..539C}. They show spectacular, asymmetric declines of up to 8 magnitudes at irregular intervals due to dust formation near the surface of the star, as well as variations due to radial oscillations at maximum light. Spectroscopically, they show similarities to hydrogen-deficient carbon stars (HdC), although the latter do not show the same declines in brightness \citep{Warner:1967lr}. Additionally, 
it has been suggested that RCBs are the evolutionary predecessors to the EHe stars,
due to strong similarities in the abundances of the two types \citep{2011MNRAS.414.3599J,Jeffery_2017}.

RCB stars are quite rare; there are currently 117 known RCB stars in the Milky Way and 30 in the Magellanic clouds \citep{2020A&A...635A..14T}.
The population of RCB stars within the Milky Way galaxy is focused near old star regions such as the bulge and the old disk. Belonging to old star regions implies that RCBs should have formed from low metallicity clouds, a conclusion that can also be drawn by the subsolar iron abundances of observed RCBs \citep{Asplund:2000qy}. 
Population synthesis models and lifetime estimates of RCB stars imply that there should be between 300 and 500 in the Milky Way \citep{2020A&A...635A..14T}.

The formation process of RCBs has long been debated \citep{Fujimoto:1977lr,1984ApJ...277..355W}, however the currently favored formation mechanism is that of a merger of two white dwarfs (WD), one carbon/oxygen (CO-) and one helium (He-). This WD merger scenario is strongly supported by an overabundance of $^{18}$O as compared to $^{16}$O, and other unusual surface abundances measured in RCB stars \citep{Clayton2007,Pandey:2008eu,Garcia-Hernandez:2010fk,2011MNRAS.414.3599J}. Modeling these merger events provides invaluable insights into the initial conditions necessary for a star to evolve into an RCB phase.
Several attempts to model the evolution from a WD merger to the RCB phase have been made 
\citep{2011ApJ...737L..34L,2013ApJ...772...59M,Menon_2018,2014MNRAS.445..660Z,2019ApJ...885...27S,2019MNRAS.488..438L}.  Most of these studies have used the 1D stellar evolution code Modules for Experiments in Stellar Astrophysics ({\it MESA}; \citealt{Paxton2011, Paxton2013, Paxton2015, Paxton2018, Paxton2019}).

\citet{2014MNRAS.445..660Z} model the WD merger using both fast and slow accretion. They find that their `destroyed disc' models, which approximate direct ingestion of the accretion disc into the envelope, can replicate the abundances shown in RCBs. Additionally, their models favor lower mass He-WDs, at masses 0.20-0.35 M$_{\sun}$. \cite{2011ApJ...737L..34L,2012A&A...542A.117L} use post processing to analyze the nucleosynthesis that occurs during a hydrodynamically simulated merger event, finding enhanced $^{18}$O and $^{19}$F, along with the production of $^{7}$Li. \cite{2013ApJ...772...59M} construct a compositional profile for use in {\it MESA} utilizing the hydrodynamic merger simulations outlined in \cite{staff2012}.  They extend this work in \cite{Menon_2018} to include subsolar metallicity, with similar results. In \cite{2019MNRAS.488..438L}, a WD merger is both simulated and evolved to the RCB phase within {\it MESA}, including a 75-isotope reaction network to find abundances in agreement with observations. Most recently, \cite{2019ApJ...885...27S} evolved a modified model of a Helium star and a spherically averaged 2D hydrodynamic merger model, both with more realistic opacities with {\it MESA}, focusing more on the structure of the RCB phase rather than the abundances. \cite{2019ApJ...885...27S} also includes a separate set of models which are inspired by WD mergers, similar to \cite{2019MNRAS.488..438L} and this work.

In this work, we build upon the models presented in \cite{2019MNRAS.488..438L} to investigate the effects of a range of initial He-burning shell temperatures and include the effects of solar and subsolar metallicities. In Section~\ref{sec:mesa}, we describe in detail the process by which we create our models within {\it MESA}, and present our results with a focus on structural differences. In Section~\ref{sec:abunds}, we present a detailed account of the models' surface abundances, including discussion of the roles of initial He-burning shell temperature and metallicity. In Section~\ref{sec:discussion}, we discuss in detail how our models differ with a focus on metallicity, production of Li, $^{18}$O overabundance, and C/O ratio, respectively. We conclude and summarize in Section~\ref{sec:conclusions}.

\section{{\it MESA} Models}
\label{sec:mesa}

In order to create our evolutionary models within {\it MESA}, we utilize the process outlined in Section 2 of \cite{2019MNRAS.488..438L}. We use {\it MESA} version r10398 with default equations of state and opacities, and focus on changes in the surface abundances produced by varying the initial temperature of the He-burning shell (hereafter T\textsubscript{He}) and the metallicity of the post-merger star. See \cite{Paxton2019} and references therein for more detailed descriptions of the MESA equations of state and opacities. In the modeling process, we aim to mimic the structure and composition of 3D hydrodynamic simulations of WD mergers from works such as \cite{staff2012,2018ApJ...862...74S}. We do this in three steps. 

First, we create the He-WD progenitor of the system using \texttt{make\_he\_wd} from the {\it MESA} test suite with a 75-isotope reaction network called \texttt{mesa\_75.net}. We assume that this He-WD will be completely disrupted in the merger process, and thus the elements will be thoroughly mixed. Under this assumption, we calculate a mass-averaged abundance profile for this He-WD and adopt it as our envelope composition profile in the merged object. For the core of the merged object, we assume that the progenitor CO-WD will be 50\% C and 50\% O. Note that this process assumes that the final composition of the post-merger object is spherically symmetric, which is not always the case. Similarly to previous studies, we do not consider dredge up from the CO-WD in the merger process, although \cite{staff2012,2018ApJ...862...74S} shows that it may be a component to be considered \citep{2013ApJ...772...59M,Menon_2018,2019MNRAS.488..438L}.

In the second step, we adjust the stellar structure of a typical star in a process we call `stellar engineering'. We begin by evolving a 0.8 M$_{\sun}$ star from pre-main sequence using the same reaction network as before, \texttt{mesa\_75.net}, until the star has a degenerate core. The mass of this star sets the mass of the post-merger object. We then stop the evolution and adjust the composition of the inner 0.55 M$_{\sun}$ or the core of the star to that of our CO-WD progenitor, and adjust the outer 0.25 M$_{\sun}$ or the envelope composition to that of our mass-averaged He-WD progenitor. After the composition is changed, we apply an entropy adjusting procedure at the base of the He-rich envelope to expand the star to a structure that mimics the temperature and density profile found at the end of our hydrodynamical merger simulations \citep{2012ApJ...748...35S,2016MNRAS.463.3461S}. An example of one such entropy adjustment can be found in Figure 1 of \cite{2019MNRAS.488..438L}. The amount of entropy injected into these models is proportional to the initial radius of the expanded object before any evolution, and inversely proportional to the initial peak temperature of the profile. This peak temperature of the initial profile is
T\textsubscript{He}, the initial temperature of the helium-burning shell of the post-merger object, and is analogous to the temperature of the ``Shell of Fire" in \cite{staff2012,2018ApJ...862...74S}. The model can be adjusted to a different total mass by beginning this step with the desired total mass, and the relative location of the constituent WDs can be adjusted during the composition and entropy adjusting procedure. 

Once we have our post-merger object engineered, we evolve our models in {\it MESA} with typical stellar astrophysics, through the RCB and EHe phases until the post-merger stars return to a WD phase. Our models assume an initial rotational velocity on the equator of 20\% of the critical Keplerian velocity, however they do not include elemental diffusion due to this rotation. All diffusion and mixing is due to traditional mixing length theory \citep{1968pss..book.....C}.

\begin{figure*}
	\includegraphics[width=\textwidth]{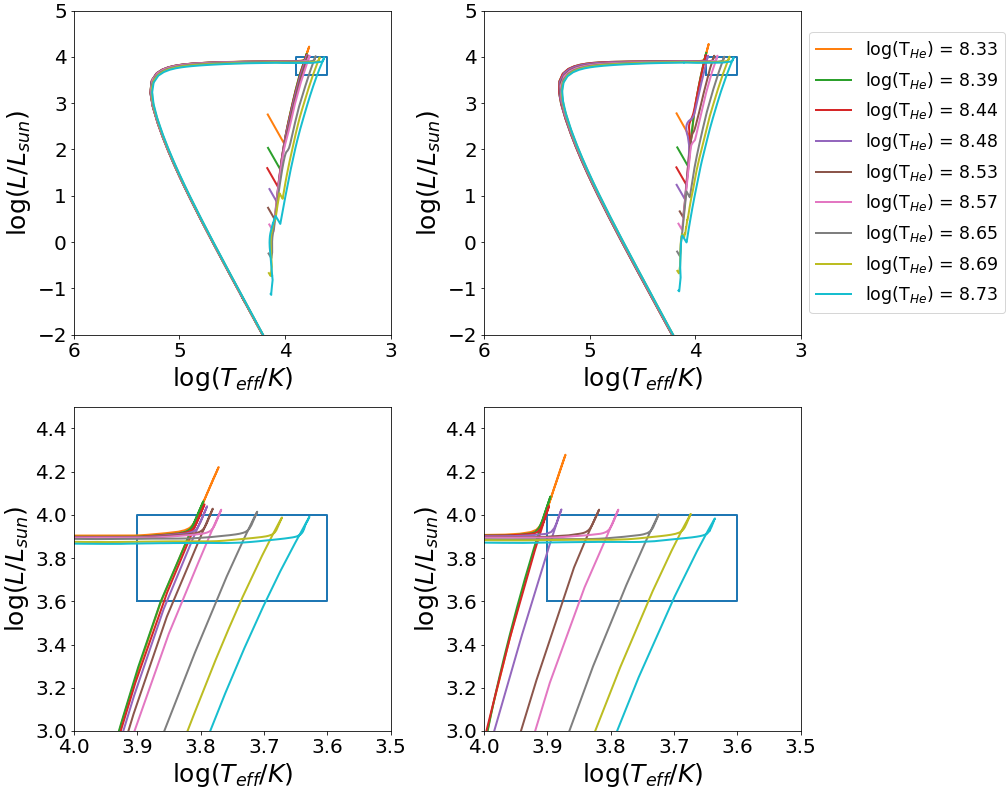}
    \caption{The path of evolution within the HR diagram for our 18 models. The left and right sides correspond to solar and subsolar metallicity, respectively. The bottom two panels are the same tracks as the top two, centered around the RCB locus. Each colored track corresponds to a different T\textsubscript{He}, labeled in the legend on the right side.}
    \label{fig:hr_panels}
\end{figure*}

\cite{Tisserand:2009fj,Tisserand:2011uq} show that most RCB stars have photospheric temperatures between 4000 and 8000 K, and have absolute V magnitudes between -5 and -3.5. Therefore, we define the region of the HR diagram with log(T\textsubscript{eff}) between 3.6 and 3.9, and log(L) between 3.6 and 4 to be the locus of the RCB stars. As seen in Figure~\ref{fig:hr_panels}, our models evolve upwards in the HR diagram to a maximum luminosity within the RCB locus, and then evolve leftwards as the surface temperature increases. We denote the time period from the point in the evolution of maximum luminosity and minimum temperature until the model evolves leftward out of the RCB locus as ``the RCB phase", and all of RCB surface abundances noted in this work are recorded at the first time step within this phase.

Using this initialization process, we created a total of 18 models, listed in Table~\ref{tab:models}. Half of these are of solar metallicity (denoted by SOL), as has been used in \cite{2019MNRAS.488..438L}, \cite{2013ApJ...772...59M}, \cite{2014MNRAS.445..660Z}, \cite{2011ApJ...737L..34L}, and \cite{2019ApJ...885...27S}. The other half of the models use a subsolar metallicity (denoted by SUB), specifically 10\% of solar values, or Z = 0.002. This is motivated both by the observed iron abundance of RCB stars, and because they reside in old star regions such as the bulge and Magellanic clouds. \cite{Menon_2018} also uses a subsolar metallicity of Z = 0.0028. Both of these values of subsolar metallicity are within the range of observed RCB abundances for Fe, -2.0 < [Fe/H] < -0.5. For each of the metallicities, we use a range of T\textsubscript{He}, the values of which are outlined in the third column of Table~\ref{tab:models}. Each model is identified with either SOL or SUB, corresponding to its metallicity, and a decimal corresponding to the log(T\textsubscript{He}).
For comparison, the range in T\textsubscript{He} used in other RCB studies is listed in Table~\ref{tab:t_he}.

\citet{2013ApJ...772...59M,Menon_2018} find that they were only able to reproduce observed RCB abundances with the inclusion of a particular mixing prescription reaching down to a precise depth, and ending before the beginning of the RCB phase of evolution. \cite{2019MNRAS.488..438L} utilize rotationally induced mixing and the default {\it MESA} mixing length theory. 
 We include rotationally induced mixing, as described in \cite{2019MNRAS.488..438L}, and thus mixing occurs naturally during the evolution of the model.

\begin{table*}
	\centering
	\caption{RCB MESA Models}
	\label{tab:models}
	\begin{tabular}{lccccccccccr} 
		\hline
		Model ID & Z & Log(T\textsubscript{He}) & RCB Radius & RCB Mass & Max L & log(T\textsubscript{eff}) & Time to RCB & RCB Lifetime & EHe Lifetime & EHe Mass\\
		 & & (K) & (log(R/R$_{\sun}$)) & (M$_{\sun}$) & log(L/L$_{\sun}$) & (K) & (10$^2$ yrs) & (10$^4$ yrs) & (10$^3$ yrs) & (M$_{\sun}$) \\
		\hline
		SOL8.33 & Solar & 8.33 & 2.09 & 0.79 & 4.22 & 3.77 & 2.1 & 1.2 & 8.3 & 0.60 \\
		SOL8.39 & Solar & 8.39 & 1.96 & 0.80 & 4.06 & 3.80 & 5.3 & 1.5 & 10.9 & 0.60 \\
		SOL8.44 & Solar & 8.44 & 1.96 & 0.80 & 4.05 & 3.79 & 5.8 & 1.7 & 9.3 & 0.60 \\
		SOL8.48 & Solar & 8.48 & 1.96 & 0.80 & 4.04 & 3.79 & 6.3 & 1.6 & 9.4 & 0.60 \\
		SOL8.53 & Solar & 8.53 & 1.98 & 0.80 & 4.03 & 3.78 & 8.2 & 1.5 & 9.0 & 0.60 \\
		SOL8.57 & Solar & 8.57 & 2.00 & 0.79 & 4.02 & 3.77 & 8.7 & 1.5 & 9.3 & 0.60 \\
		SOL8.65 & Solar & 8.65 & 2.11 & 0.79 & 4.01 & 3.71 & 9.1 & 1.2 & 6.7 & 0.59 \\
		SOL8.69 & Solar & 8.69 & 2.17 & 0.79 & 3.99 & 3.67 & 9.0 & 1.2 & 6.8 & 0.59 \\
		SOL8.73 & Solar & 8.73 & 2.26 & 0.79 & 3.99 & 3.63 & 8.3 & 1.0 & 6.1 & 0.58 \\
		\hline

		SUB8.33 & Subsolar & 8.33 & 1.92 & 0.79 & 4.28 & 3.87 & 1.5 & 1.9 & 14.4 & 0.61 \\
		SUB8.39 & Subsolar & 8.39 & 1.78 & 0.80 & 4.08 & 3.90 & 4.6 & 2.6 & 18.7 & 0.62 \\
		SUB8.44 & Subsolar & 8.44 & 1.75 & 0.80 & 4.04 & 3.90 & 7.0 & 2.7 & 19.3 & 0.62 \\
		SUB8.48 & Subsolar & 8.48 & 1.75 & 0.80 & 4.02 & 3.88 & 8.6 & 2.4 & 16.1 & 0.62 \\
		SUB8.53 & Subsolar & 8.53 & 1.90 & 0.80 & 4.02 & 3.82 & 8.7 & 1.9 & 12.0 & 0.61 \\
		SUB8.57 & Subsolar & 8.57 & 1.96 & 0.80 & 4.02 & 3.79 & 8.6 & 1.6 & 9.8 & 0.60 \\
		SUB8.65 & Subsolar & 8.65 & 2.08 & 0.80 & 4.00 & 3.72 & 9.4 & 1.3 & 7.0 & 0.59 \\
		SUB8.69 & Subsolar & 8.69 & 2.18 & 0.79 & 4.00 & 3.67 & 9.2 & 1.1 & 5.6 & 0.59 \\
		SUB8.73 & Subsolar & 8.73 & 2.24 & 0.80 & 3.98 & 3.64 & 8.7 & 1.0 & 5.5 & 0.58 \\
		\hline
	\end{tabular}
\end{table*}

\begin{table}
	\centering
	\caption{ T\textsubscript{He} in Previous Studies.}
	\label{tab:t_he}
	\begin{tabular}{lr} 
		\hline
		Work & log(T\textsubscript{He})\\
		\hline
		\cite{Clayton2007} & 8.22\\ 
		\cite{2019MNRAS.488..438L} & $\sim$8.45 - 8.70\\ 
		\cite{2014MNRAS.445..660Z} (Slow Accretion) & $\sim$8.29 - 8.35\\ 
		\cite{2014MNRAS.445..660Z} (Fast Accretion) & $\sim$8.40 - 8.45\\ 
		\cite{Menon_2018} & 8.08 and 8.40\\ 
		\cite{2013ApJ...772...59M} & 8.11 and 8.38\\ 
		\cite{2019ApJ...885...27S} & $\sim$8.2 - 8.4\\ 
		\cite{2011ApJ...737L..34L} & $\sim$8.5\\ 
		Munson et al. (in preparation) & $\sim$8.38\\ 
		\cite{2018ApJ...862...74S} & $\sim$7.84 - 8.4\\ 
		\hline
	\end{tabular}
\end{table}

\begin{figure}
	\includegraphics[width=\columnwidth]{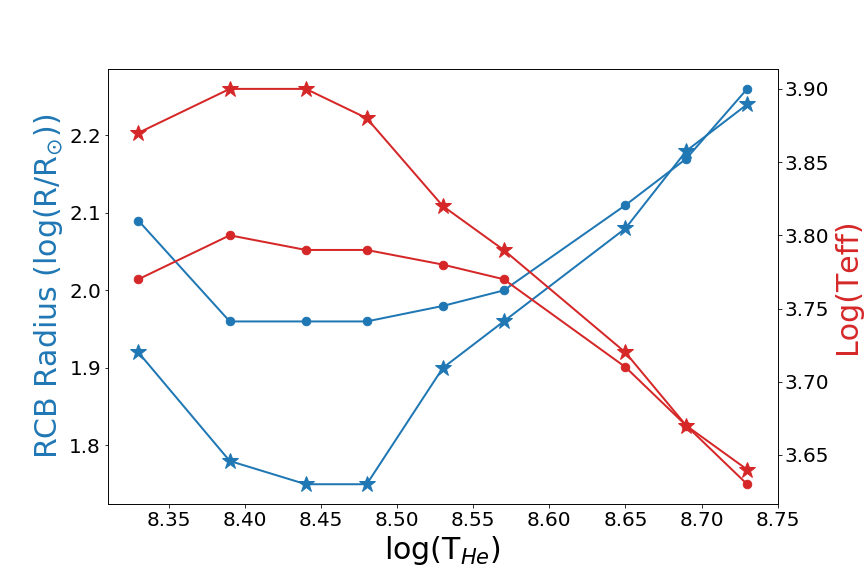}
    \caption{The relationship between radius and T\textsubscript{eff} in the RCB phase as a function of the He-burning shell temperature, T\textsubscript{He}. Points marked with a circle indicate solar metallicity, and those marked with a star indicate subsolar metallicity. Radius and effective temperature trends are denoted by the colors blue and red, respectively.}
    \label{fig:teff_rad}
\end{figure}

Figure~\ref{fig:hr_panels} shows the evolution of our models within the HR diagram. The evolution begins with a short relaxation within {\it MESA} before the luminosity increases at approximately constant surface temperature, until reaching the RCB locus on a timescale of $\sim$10$^2$ years. The star remains at this peak luminosity, minimum temperature phase for $\sim$10$^4$ years, after which it moves quickly leftwards, leaving the RCB locus, and entering the region where the EHe stars reside. The star passes through this EHe region in $\sim$10$^3$ years, rapidly shrinking due to mass loss until it reaches degeneracy and turns onto the WD cooling track. A notable result is that our models with lower T\textsubscript{He} produce RCBs with higher T\textsubscript{eff}. This effect occurs because the high temperatures within the He-burning shell drive increased energy generation, pushing the radius of the star further outward, and reducing T\textsubscript{eff}. This can be confirmed in Figure~\ref{fig:teff_rad} as we see that the T\textsubscript{eff} and the radius of the RCB phase are inversely correlated. Another clear evolutionary effect, most evident in Figure~\ref{fig:hr_panels}, is that the models with the lowest T\textsubscript{He} within each metallicity subset congregate around a single T\textsubscript{eff}. For solar models this is log(T\textsubscript{eff}) $\sim$ 3.8 and for subsolar models this is log(T\textsubscript{eff}) $\sim$ 3.9. Conversely, at large T\textsubscript{He} the two metallicity subsets are indistinguishable from each other, and instead their radius and temperature depends only on T\textsubscript{He}, rather than metallicity (best seen in Figure~\ref{fig:teff_rad}). 

\begin{figure}
	\includegraphics[width=\columnwidth]{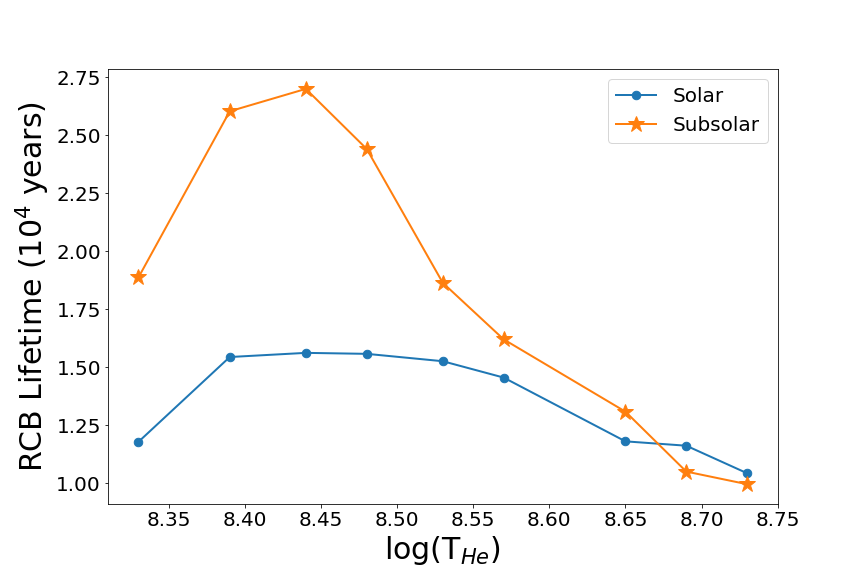}
    \caption{The relationship between T\textsubscript{He} and the lifetime of the RCB phase. The solar models are marked by blue filled circles, and the subsolar models are marked by orange stars. The RCB phase lifetime is in units of 10$^4$ years.}
    \label{fig:rcblifetime}
\end{figure}

Additionally, we calculate the time that each model spends in the RCB phase, shown in Figure~\ref{fig:rcblifetime}. This is calculated by first locating the beginning of the RCB phase, which is where the star has its peak luminosity and minimum T\textsubscript{eff}. Then, we calculate the age where the star exits the left side of the RCB locus. The difference between these two times is the RCB lifetime. We find that this lifetime is largely dependent on the rate of mass loss going on in the star at this point of the evolution. Our models spend on the order of 10$^4$ years in the RCB phase, see Table~\ref{tab:models}. The subsolar models spend more time in this phase on average, which we expect as lower metallicity stars exhibit slower mass loss \citep{1992ApJ...401..596L,2000ASPC..204..395L} and thus spend more time in RCB phase before evolving leftwards. We note, however, that the Bl{\"o}cker wind prescription (Eqn~\ref{eqn:blocker}, \citep{1995A&A...297..727B}) does not explicitly contain metallicity as a parameter. Note that this form of the Bl{\"o}cker mass loss has been simplified in order to emphasize the scaling relationships due to mass, radius, and luminosity.
\begin{equation}
\label{eqn:blocker}
    \dot{M_{B}} = 1.932 * 10^{-21} \eta L^{3.7} R M^{-3.1}
\end{equation}
The subsolar metallicity models have smaller radii due to their lower opacity, and although the radius has a weaker scaling than luminosity, the large decrease in radius dominates and decreases the mass loss in spite of slightly higher luminosities.
We also see that the curve has an overall negative slope; stars with a higher T\textsubscript{He} spend less time in the RCB phase, and evolve more quickly. Similarly to the metallicity case, the models with higher T\textsubscript{He} have larger radii, and slightly lower luminosity, however the increase in radius dominates and increases the mass loss for these models.

As mentioned, the RCB lifetime we calculate is largely dependent on the adopted mass loss prescription. Our models utilize a typical Bl{\"o}cker AGB wind prescription with efficiency parameter $\eta$ = 0.075 \citep{1995A&A...297..727B}. See \cite{2019ApJ...885...27S} for an excellent discussion on the effects of varying $\eta$. Table~\ref{tab:eta} lists the values of $\eta$ used in previous RCB modeling. To investigate the effects of efficient winds on our models, and to confirm that the lifetime is dependent on mass loss, we ran our SOL8.39, SOL8.69, SUB8.39, and SUB8.69 models with $\eta$ = 0.005 as was used in \citep{2019MNRAS.488..438L}. This is an order of magnitude less efficient mass loss than is used in the models presented here. The evolution of surface abundances is not affected by the value of $\eta$. The main effects of changing $\eta$ are the lifetime of the star in the RCB locus and the final mass of the RCB star. These reduced $\eta$ models have significantly longer RCB and EHe lifetimes, on the order of 10$^5$ and 10$^4$, respectively, and are more massive when they leave the RCB locus and become EHe stars. This information is summarized in Table~\ref{tab:etamodels}.

\begin{table}
	\centering
	\caption{$\eta$ Values for Bl{\"o}cker AGB winds.}
	\label{tab:eta}
	\begin{tabular}{lr} 
	    \hline
		Work & $\eta$\\
		\hline
		This Work & 0.075\\
		\cite{2019MNRAS.488..438L} & 0.005\\ 
		\cite{2014MNRAS.445..660Z} & 0.02 and 0.1\\ 
		\cite{Menon_2018} & 0.05\\ 
		\cite{2019ApJ...885...27S} &  0, 0.01, 0.02, and 0.05\\ 
		\hline
	\end{tabular}
\end{table}

\begin{table}
    \centering
    \caption{Models with $\eta$ = 0.005}
    \begin{tabular}{lccc}
        \hline
         Model ID & EHe Mass & RCB lifetime & EHe lifetime  \\
          & (M$_{\sun}$) & (10$^5$ yr) & (10$^4$ yr) \\
         \hline
         SOL8.39 & 0.70 & 1.4 & 1.7 \\
         SOL8.69 & 0.68 & 1.3 & 1.4 \\
         SUB8.39 & 0.72 & 1.6 & 1.9 \\
         SUB8.69 & 0.67 & 1.3 & 1.4 \\ 
    \end{tabular}
    \label{tab:etamodels}
\end{table}

\section{Surface Abundances}
\label{sec:abunds}

All surface abundances in this paper are calculated using the typical expression,
\begin{equation}
    \log\epsilon(X) = \log(X) - \log(\mu_X) + 12.15
\end{equation}
where X represents the surface mass fraction of the element and $\mu_X$ is the mean atomic mass of that element. We define the surface zones where we measure the abundances to be those with optical depth $<$ 1. This value represents the log of the number of ions of element X in a sample that is assumed to contain a total of 10$^{12.15}$ ions. A detailed account of all tabulated surface abundances is included in Table~\ref{tab:abunds}.

\begin{figure*}
	\includegraphics[width=\textwidth]{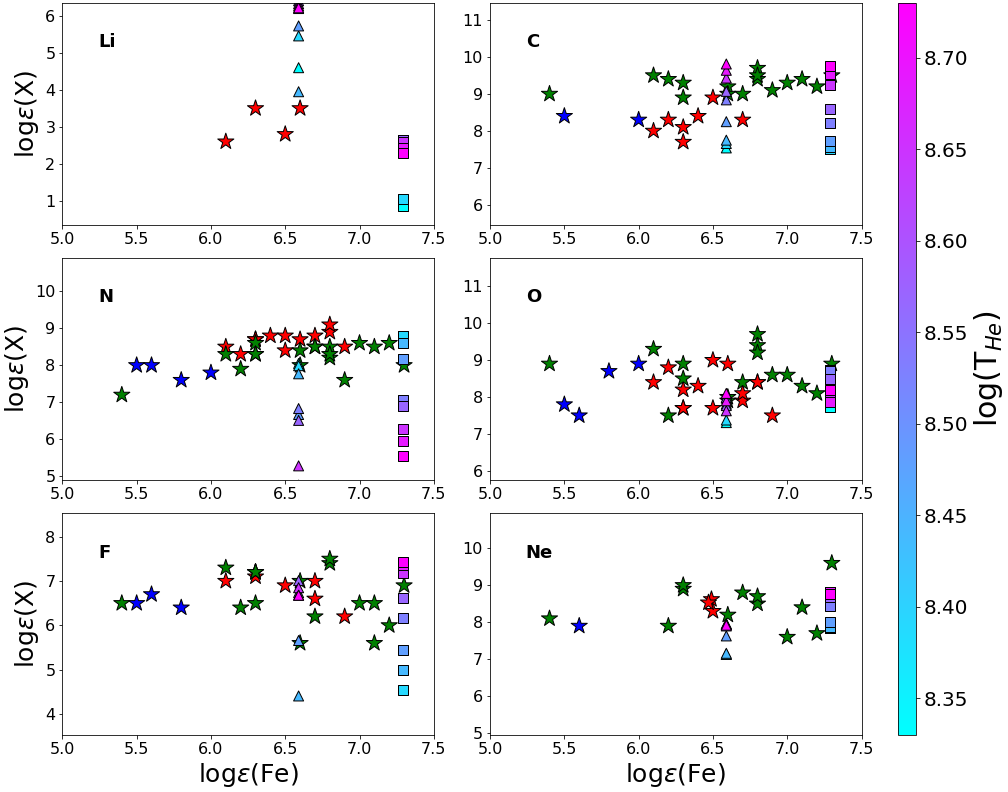}
    \caption{The observed surface abundances from of known majority RCBs, minority RCBs , and EHe stars are marked by red, blue, and green stars, respectively. Our SOL and SUB models are marked by colored squares and triangles, respectively, where the color indicates the log(T\textsubscript{He}) for that model according to the color bar on the right.}
    \label{fig:all_abunds}
\end{figure*}

\begin{figure}
	\includegraphics[width=\columnwidth]{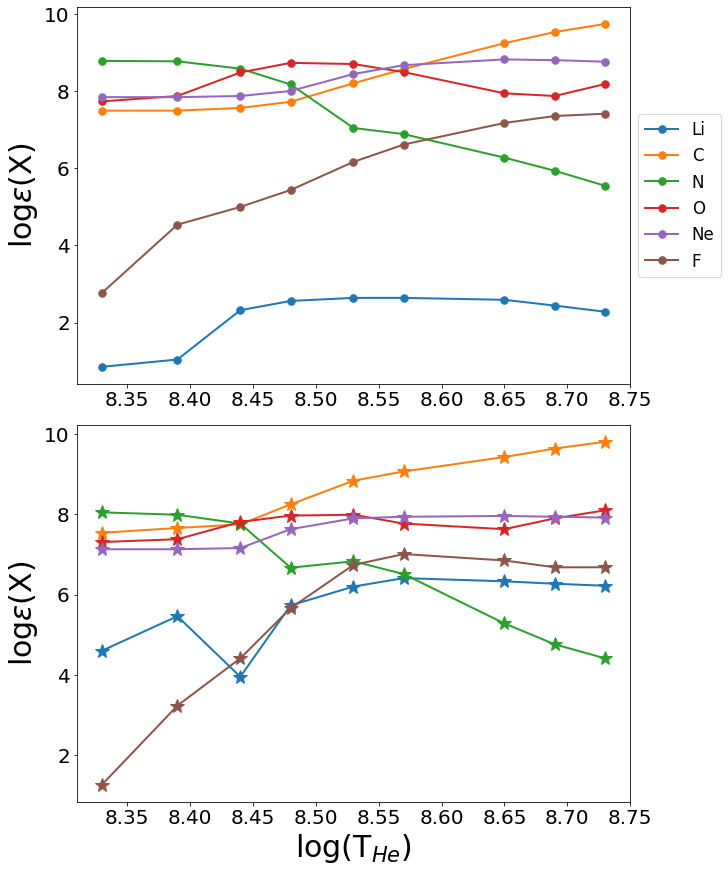}
    \caption{The logarithmic abundances by number as a function of T\textsubscript{He}. The upper panel contains solar metallicity models, and the lower panel contains subsolar metallicity models. Each colored line represents a different element according to the legend on the right.}
    \label{fig:licnonef}
\end{figure}

We compare the surface abundances from our models  to those measured from observations. The bulk of these values for the observed RCBs and EHe stars comes from \cite{2011MNRAS.414.3599J}, which aggregates data from a range of works such as \cite{Asplund:2000qy} and \cite{Pandey:2008eu}. We also compare to the abundances calculated for the Sun \citep{Lodders_2003}. A complete representation of our models' surface abundances compared to the observations is presented in Figure~\ref{fig:all_abunds}. In Figure~\ref{fig:licnonef} we present the trend of surface abundances for a select few elements as a function of T\textsubscript{He} for both metallicities. 

The formation of RCB stars by the merger of a CO- and a He-WD binary is now well supported. Previous models, listed in Table~\ref{tab:t_he}, can account for the main observed abundance peculiarities seen in the RCB stars, namely the $^{16}$O/$^{18}$O and $^{12}$C/$^{13}$C ratios, F
abundances, the C, N, O, Ne, and s-process abundances. A mixture of the products from H- and He-burning are needed, in particular, to account for the high N abundance, implying an RCB's surface exhibits CNO-cycled material. The initial temperature of the He-burning shell (T\textsubscript{He}) is of critical importance, as when T\textsubscript{He} increases above 2 $\times 10^8$ K, the N abundance decreases to below RCB star levels. 

The key reactions which lead to the observed RCB star surface abundances are
$^{13}$C($\alpha$,n)$^{16}$O
and
$^{14}$N($\alpha$,$\gamma$)$^{18}$F($\beta^+$)$^{18}$O($\alpha$,$\gamma$)$^{22}$Ne.\\
The former reaction burns away nearly all of the $^{13}$C, increasing the $^{12}$C/$^{13}$C ratio and providing neutrons for s-processing. If the latter reaction proceeds partially, it leads to a large increase in $^{18}$O although the isotope will eventually be fully converted into $^{22}$Ne if the reaction is allowed to proceed to completion. Therefore, the mixing of this partially burned material, $^{18}$O, to the surface must happen fairly early in the evolution of the RCB star. This second reaction also causes the O abundance to increase while the triple-$\alpha$ reaction slowly increases $^{12}$C. T\textsubscript{He} is generally not high enough for $^{12}$C($\alpha$,$\gamma$)$^{16}$O to proceed strongly.
Large amounts of $^{19}$F are also created by
$^{14}$N(n,p)$^{14}$C(p,$\gamma$)$^{15}$N($\alpha$,$\gamma$)$^{19}$F
and
$^{18}$O(p,$\gamma$)$^{19}$F. However, our reaction network does not contain the isotope $^{14}$C, and thus we are unable to track the neutron poison reaction $^{14}$N(n,p)$^{14}$C. Additionally, if there is a remnant  H-shell around the CO-WD, then that hydrogen will get mixed into the envelope during the merger, allowing p-capture reactions to proceed. This can have the effect of increasing $^{16}$O/$^{18}$O and decreasing $^{12}$C/$^{13}$C if they become too extreme \citep{Clayton2007}.

\begin{table*}
	\centering
	\caption{Model Abundances. RCB majority abundances are taken from \protect\cite{Asplund:2000qy}, and Solar values are taken from \protect\cite{Lodders_2003}.}
	\label{tab:abunds}
	\begin{tabular}{lccccccccr} 
		\hline
		Model ID & Li & C & $^{12}$C/$^{13}$C & N & O & $^{16}$O/$^{18}$O & C/O & F & Ne \\
		\hline
		SOL8.33 & 0.85 & 7.5 & 15 & 8.79 & 7.74 & 1.4 x 10$^{6}$ & 0.58 & 2.77 & 7.85\\
		SOL8.39 & 1.04 & 7.5 & 2.1 x 10$^2$ & 8.78 & 7.88 & 3.24 & 0.42 & 4.54 & 7.85\\
		SOL8.44 & 2.32 & 7.57 & 5.5 x 10$^{7}$ & 8.59 & 8.49 & 0.25 & 0.12 & 5.0 & 7.88\\
		SOL8.48 & 2.56 & 7.73 & 1.6 x 10$^{8}$ & 8.18 & 8.74 & 0.13 & 0.09 & 5.44 & 8.01\\
		SOL8.53 & 2.64 & 8.21 & 1.0 x 10$^{8}$ & 7.05 & 8.71 & 0.14 & 0.32 & 6.17 & 8.45\\
		SOL8.57 & 2.64 & 8.58 & 1.4 x 10$^{8}$ & 6.89 & 8.5 & 0.23 & 1.20 & 6.62 & 8.68\\
		SOL8.65 & 2.59 & 9.25 & 1.8 x 10$^{9}$ & 6.28 & 7.95 & 2.88 & 19.95 & 7.18 & 8.83\\
		SOL8.69 & 2.44 & 9.54 & 6.9 x 10$^{9}$ & 5.94 & 7.88 & 13.8 & 45.71 & 7.36 & 8.81\\
		SOL8.73 & 2.28 & 9.75 & 1.3 x 10$^{10}$ & 5.55 & 8.19 & 102 & 36.31 & 7.42 & 8.77\\
		\hline

		SUB8.33 & 4.6 & 7.54 & 427 & 8.05 & 7.31 & 1.79 x 10$^{6}$ & 1.70 & 1.26 & 7.13\\
		SUB8.39 & 5.46 & 7.66 & 8.32 & 7.99 & 7.38 & 1.1 x 10$^{3}$ & 1.91 & 3.23 & 7.13\\
		SUB8.44 & 3.95 & 7.75 & 2.88 x 10$^{7}$ & 7.77 & 7.81 & 1.07 & 0.87 & 4.41 & 7.16\\
		SUB8.48 & 5.73 & 8.25 & 8.51 x 10$^{8}$ & 6.67 & 7.97 & 0.58 & 1.91 & 5.66 & 7.63\\
		SUB8.53 & 6.2 & 8.84 & 9.0 x 10$^{9}$ & 6.83 & 7.99 & 5.5 & 7.08 & 6.74 & 7.9\\
		SUB8.57 & 6.41 & 9.07 & 1.62 x 10$^{9}$ & 6.51 & 7.77 & 11.5 & 19.96 & 7.01 & 7.94\\
		SUB8.65 & 6.33 & 9.43 & 1.78 x 10$^{10}$ & 5.28 & 7.63 & 66.1 & 63.10 & 6.85 & 7.96\\
		SUB8.69 & 6.27 & 9.64 & 2.24 x 10$^{10}$ & 4.76 & 7.9 & 309 & 54.95 & 6.68 & 7.94\\
		SUB8.73 & 6.22 & 9.81 & 2.82 x 10$^{10}$ & 4.41 & 8.1 & 1.1 x 10$^3$ & 51.29 & 6.68 & 7.92\\
		\hline
		RCB Majority&2.6--3.5&7.7--8.9&$>$500&8.3--9.1&7.5--9.0&$\sim$1&$\sim$1&6.9--7.2&7.9--8.9\\
		Sun&1.1&8.4&89&7.8&8.7&500&0.5&4.5&7.9\\
		\hline
	\end{tabular}
\end{table*}

\subsection{Carbon, Nitrogen, and Oxygen}
\label{sec:CNO}

Carbon, nitrogen, and oxygen are distinctly linked together in the CNO-cycle.
The dominant product of complete H-burning via the CNO-cycle is N, since $^{14}$N has the smallest nuclear $p$-capture cross section of the stable CNO elements \citep{Clayton2007}. However in RCB stars, C is the most abundant with N second and O third. 

Carbon is the primary source of opacity in RCB atmospheres. Its abundance has been difficult to measure directly because of saturated CI lines in the spectra. Therefore, previous studies typically made an assumption as to the value of the C/He ratio, leading to the discrepancy between model atmosphere predictions and observations known as ``the carbon problem" \citep{Asplund:2000qy}. We compare our models to the more accurate abundances directly measured using C$_2$ bands  for RCB stars \citep{2012ApJ...747..102H}, and those from \cite{2011MNRAS.414.3599J} for EHe stars. For discussion regarding the abundance of the isotope $^{13}$C, see Section~\ref{sec:c13}.

Nearly all RCB stars show enriched N abundances relative to solar, with the majority RCBs having an average abundance of 8.65, 0.75 dex higher than the solar value. This is curious considering the lack of H in these stars, and thus the unavailability of H replenishment throughout the CNO cycle. Further, the majority RCBs in our sample have an average [N/Fe] = 1.7, which is higher than what can be achieved solely through CNO-cycling, therefore there must be some contamination due to He-burning products \citep{Asplund:2000qy}. The dominant reaction that destroys N is $^{14}$N($\alpha$,$\gamma$)$^{18}$F($\beta^+$)$^{18}$O, which produces an element integral to the identification of RCB stars (see Section~\ref{sec:o18}), and is responsible for a decrease the N abundance.

Oxygen is the least abundant CNO element in RCB stars. The $\alpha$-capture reaction with the largest cross section, and thus the first reaction to occur at the onset of He-burning, is $^{13}$C($\alpha$,n)$^{16}$O. Thus, $^{16}$O is quickly enhanced. Additionally, we identified another important reaction chain, $^{14}$N($\alpha$,$\gamma$)$^{18}$F($\beta^+$)$^{18}$O($\alpha$,$\gamma$)$^{22}$Ne. Both of these important $\alpha$-captures, as well as the CNO-cycle, play a large role in the observed abundances of O in RCBs. In addition to being the least abundant CNO element, O also exhibits the largest spread in observed abundances among the RCB stars, spanning a range of 1.5 dex in the majority sample alone, whereas C spans 1.2 dex and N spans 0.8 dex. The RCBs have an average abundance of $\log\epsilon$(O\textsubscript{RCB}) = 8.2, ranging from 7.5 to 9.0, with the EHes being slightly more abundant at $\log\epsilon$(O\textsubscript{EHe}) = 8.6, ranging from 7.5 to 9.7 \citep{2011MNRAS.414.3599J}. The reason for this large spread in observations is not easily explained. 

\begin{figure}
	\includegraphics[width=\columnwidth]{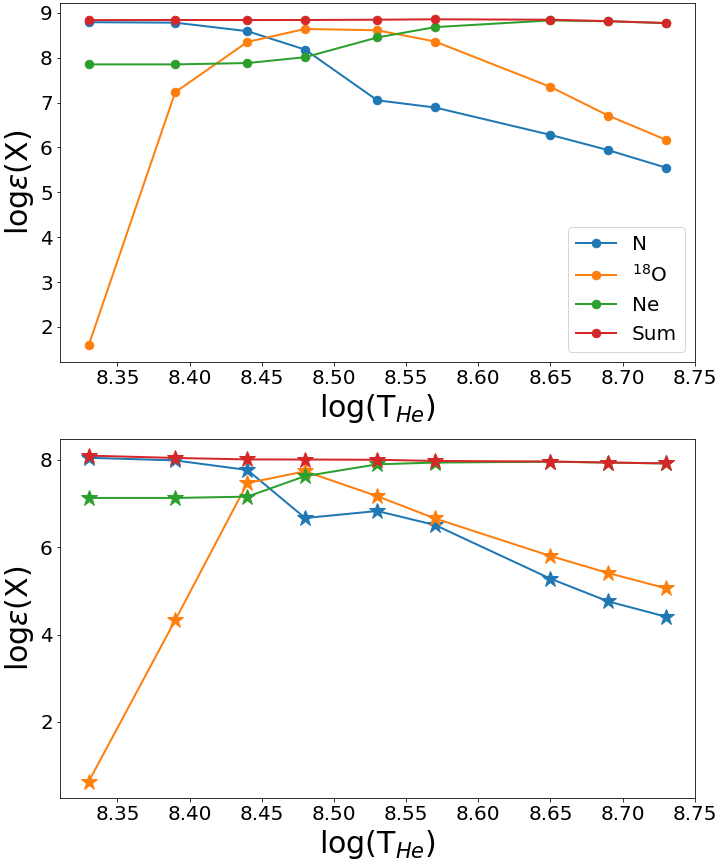}
    \caption{The blue, orange, and green lines track the logarithmic abundance log$\epsilon$(X) for N, $^{18}$O, and Ne in our models as a function of T\textsubscript{He}. The red line shows the sum of these three elements. The upper panel contains solar metallicity models, and the lower panel contains subsolar metallicity models.}
    \label{fig:N-O18-Ne-sum}
\end{figure}

In Figures~\ref{fig:all_abunds} and \ref{fig:licnonef}, we present the CNO abundances of our models. Our models show a monotonic increase in C abundance with respect to T\textsubscript{He}, whereas N is the only element to show a monotonic decrease with respect to T\textsubscript{He}. O abundance is mostly steady with T\textsubscript{He}, with slight oscillations from the mean. It is clear that the C abundance does not depend strongly on the metallicity of the RCB, as the range of calculated values is nearly the same for the SOL and SUB models. Only our coldest models lie within the range of N observations. The reduction of N as T\textsubscript{He} increases can be traced in Figure~\ref{fig:N-O18-Ne-sum}, which shows the dependence of the conversion of $^{14}$N into $^{18}$O on T\textsubscript{He}. We can see clearly that the abundance of N is primarily dependent on the temperature of the He-burning region and by extension the strength of the $\alpha$-captures on $^{14}$N. Therefore, in order for our RCB models to retain their enrichment in N, they must have a He-burning shell initially cool enough to prevent its rapid destruction. Since the spread in observed O abundances is so large, it appears that nearly all of our models fit within the observations, the exceptions being the two coldest subsolar models. The relative agreement of our models with observations for these three elements is marked in Table~\ref{tab:agree}.

Combining the C abundances for 11 RCB stars derived from the C$_2$ bands with measured O abundances for the same stars, we find that C/O$\sim$1 \citep{Asplund:2000qy,2012ApJ...747..102H}. The individual ratios range from 0.5 to 3.98. It is likely that C/O$>$1 for all of the RCB stars. In the spectra of the stars cool enough to display molecular bands, CO is the only oxygen molecule seen along with C$_2$ and CN bands. Also, C$_{60}$ has been detected in DY Cen and possibly V854 Cen \citep{2011ApJ...729..126G}. No molecules such as TiO, seen in cool stars with C$<$O, are detected in any RCB star. In addition, there is strong evidence that the dust forming around RCB stars is entirely amorphous carbon. The IR spectral continuum is featureless and there is no sign of silicate features. So it is likely that after the C and O combine in the cool RCB stars, there is leftover C to make carbon molecules and dust. 

\begin{figure}
	\includegraphics[width=\columnwidth]{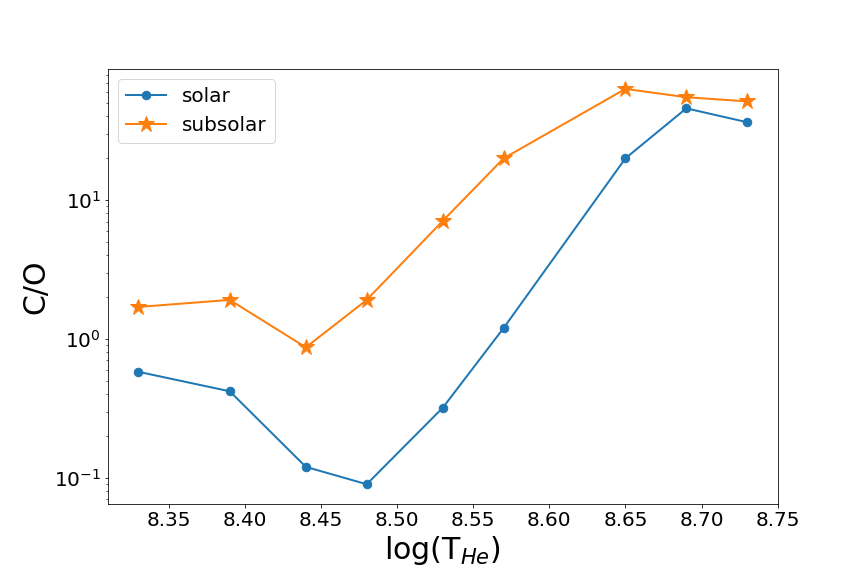}
    \caption{C/O ratio vs T\textsubscript{He} for our SOL models (blue closed circles) and our SUB models (orange stars).}
    \label{fig:CO_temp}
\end{figure}

Figure~\ref{fig:CO_temp} shows that the C/O ratios in our models increases dramatically at high temperatures. At these kinds of temperatures, the triple-$\alpha$ reaction is occurring at a very high rate, but the models are not hot enough to efficiently convert that C into O through $\alpha$-capture. Our cooler solar models have C/O ratios that are much smaller than one, whereas the cooler subsolar models are all very near C/O = 1, the desired region.

\subsection{Neon}
\label{sec:neon}
 There are only four RCBs with measured Ne abundances: Y Mus, V3795 Sgr, ASAS-RCB-8, and V532 Oph \citep{Asplund:2000qy,Hema_2017}. These four RCBs have Ne abundances ranging from log$\epsilon$(Ne) = 7.9 to 8.6, which are at or slightly above the Solar value of 7.95.
 The EHe stars are, in general, appreciably overabundant in Ne \citep{2020ApJ...891...40B}.
 The EHes range from log$\epsilon$(Ne) = 7.6 to 9.6. 
As discussed above, $^{22}$Ne is the resultant element from the important reaction chain from $^{14}$N to $^{18}$O to $^{22}$Ne. It also can 
be a source of neutrons at very high temperatures due to the reaction $^{22}$Ne($\alpha$,n)$^{25}$Mg.

Figure~\ref{fig:licnonef} shows that as T\textsubscript{He} increases, the abundance of Ne also increases, although very slightly. Where our models fit in with the observations is shown in Figure~\ref{fig:all_abunds}. Unsurprisingly, the subsolar models have an appreciably smaller abundance of Ne, by about 1 dex, since the N abundance is similarly lower. The models which agree with the observations are marked in Table~\ref{tab:agree}. Upon further inspection of these models, we find that in the cooler models the most abundant isotope of Ne is $^{20}$Ne, as is typical. However, as we move to the hotter models, the most abundant isotope is $^{22}$Ne. This follows from the total conversion of $^{18}$O into $^{22}$Ne occurring at higher temperatures, whereas at lower temperatures the dominant source of Ne is $^{16}$O($\alpha$,$\gamma$)$^{20}$Ne. Both of these isotopes,$^{20}$Ne and $^{22}$Ne, can undergo another $\alpha$-capture to produce $^{24}$Mg + $\gamma$ and $^{25}$Mg + neutron, respectively. The abundances of these two isotopes of Mg track well with the respective Ne isotopes.

\subsection{\textsuperscript{13}C}
\label{sec:c13}

The $^{12}$C/$^{13}$C ratio is one of the key methods of distinguishing between RCB stars and carbon stars. In general, RCB stars have no detectable $^{13}$C and $^{12}$C/$^{13}$C $\ga$ 100 while in cool carbon stars the ratio is typically $<$ 100
\citep{1977PASJ...29..711F}.
 However, a few RCB stars do have detectable $^{13}$C. V CrA, V854 Cen, VZ Sgr, and UX Ant have measured $^{12}$C/$^{13}$C  $\la$ 25 \citep{2008MNRAS.384..477R,Hema_2017}.  
 
$^{13}$C also plays a critical role in the nucleosynthesis of RCB stars where it is the first $\alpha$-capture reaction to occur at the onset of He-burning, $^{13}$C($\alpha$,n)$^{16}$O, is the dominant source of neutrons used in synthesizing $s$-process elements, which are enhanced in RCB stars \citep{Clayton2007}. As the $^{13}$C neutron reaction progresses, the  $^{12}$C/$^{13}$C ratio will begin to increase dramatically as the $^{13}$C abundance drops to nearly zero. Most RCBs show no $^{13}$C features in their spectra \citep{2020A&A...635A..14T}, making it hard to have an exact estimate of $^{12}$C/$^{13}$C. 

\begin{figure}
	\includegraphics[width=\columnwidth]{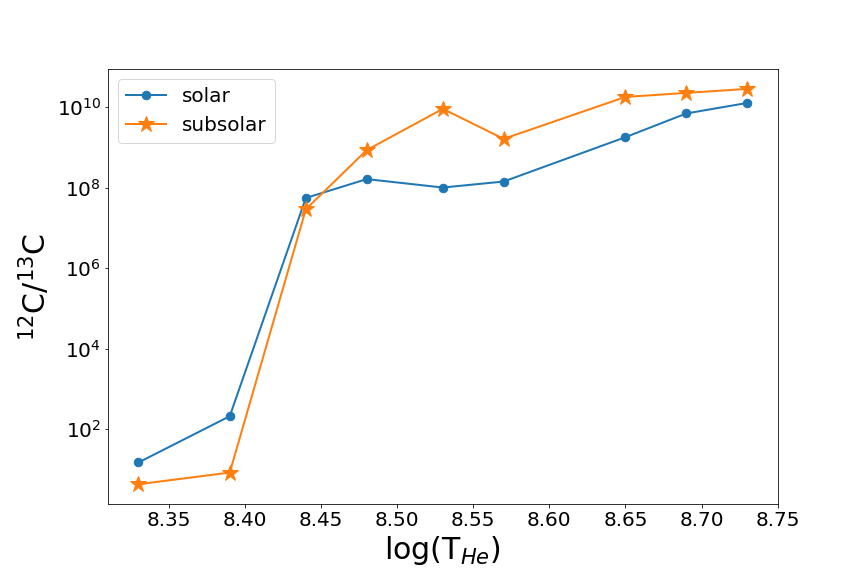}
    \caption{$^{12}$C/$^{13}$C as a function of T\textsubscript{He} for solar (blue closed circles) and subsolar (orange stars) models.}
    \label{fig:c12c13}
\end{figure}

Figure~\ref{fig:c12c13} shows the trend of $^{12}$C/$^{13}$C ratio as a function of T\textsubscript{He} for our models. Our two coldest models for both metallicities have much smaller values of this ratio than our other models, which have nearly zero $^{13}$C. As known RCBs have quite large $^{12}$C/$^{13}$C ratios, this constrains us to favor the warmer models, those with log(T\textsubscript{He}) $>$ 8.40, marked in Table~\ref{tab:agree}.

\subsection{\textsuperscript{18}O and \textsuperscript{19}F}
\label{sec:o18}

The large overabundance of $^{18}$O and $^{19}$F measured in RCB stars is unique to these stars and therefore key to their identification. 
\cite{Warner:1967lr} predicted that RCB stars would be enriched in $^{18}$O, and 40 years later it was confirmed when it was discovered that RCB stars with measurable CO bands show greatly enhanced $^{18}$O relative to $^{16}$O. \citep{2005ApJ...623L.141C,Clayton2007}. Most RCB stars exhibit ratios of $^{16}$O/$^{18}$O on the order of unity, as opposed to the solar value of 500. 
$^{19}$F is enhanced 800 to 8000 times compared to solar values \citep{Pandey:2008eu,2020ApJ...891...40B}.

While EHe stars are not cool enough to exhibit the CO bands necessary to measure the oxygen isotopic ratio, they do show enhanced $^{19}$F, cementing the proposed close evolutionary relationship with the RCB stars. The enhancement of $^{18}$O and $^{19}$F has been the most compelling evidence for the WD-merger formation, rather than a final helium shell flash, since the latter contains temperatures that would convert $^{14}$N completely into $^{22}$Ne, rather than stopping in the middle at $^{18}$O \citep{Clayton2007}. 

\begin{figure}
	\includegraphics[width=\columnwidth]{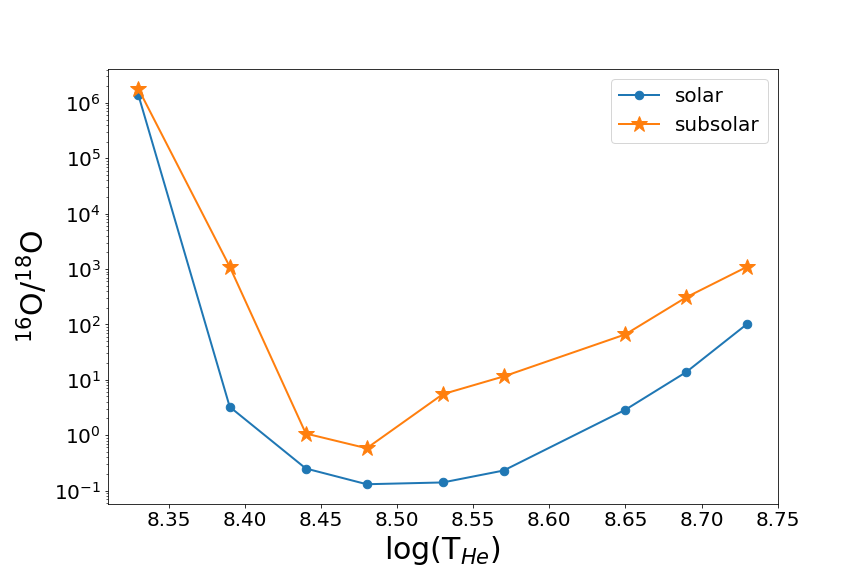}
    \caption{$^{16}$O/$^{18}$O vs T\textsubscript{He} for our SOL models (blue closed circles) and our SUB models (orange stars).}
    \label{fig:O16O18_temp}
\end{figure}

Figure~\ref{fig:O16O18_temp} shows the dependence of the O isotopic ratio on T\textsubscript{He}. This curve follows closely to the inverse of the $^{18}$O curve from Figure~\ref{fig:N-O18-Ne-sum}, since the reaction $^{14}$N($\alpha$,$\gamma$)$^{18}$F($\beta^+$)$^{18}$O($\alpha$,$\gamma$)$^{22}$Ne is what controls the abundance of this isotope. The abundance of $^{22}$Ne increases as the abundances of $^{14}$N and $^{18}$O decrease.

Figure~\ref{fig:licnonef} shows the trend of F abundance as a function of T\textsubscript{He}, and Figure~\ref{fig:all_abunds} shows where our models fit into the observations. It is clear that the F enhancement occurs at higher T\textsubscript{He}, thus favoring a higher temperature model.  As T\textsubscript{He} increases in our models, the $\alpha$-capture reactions happen more rapidly, allowing both the creation of more $^{18}$O that will then $p$-capture to $^{19}$F, and the direct creation of F from $^{15}$N. However, the trend plateaus at the highest temperatures where $^{18}$O is preferentially converted into $^{22}$Ne by $\alpha$-capture. The models that lie within the observed range of values are marked in Table~\ref{tab:agree}.

\subsection{Lithium}
\label{sec:lithium}

Of the known RCB stars, only four show the presence of Li, including the eponymous R CrB itself \citep{Asplund:2000qy}. \cite{Clayton2007} asserted that the production of Li was very hard to explain in a WD-merger scenario, especially in combination with an enriched $^{18}$O environment. However, \cite{2012A&A...542A.117L} posits that the observed Li abundance is related to the viewing angle of the RCB star, since a true post-merger object will not be spherical. They suspect that the Li is formed and transported to the surface through the Cameron-Fowler mechanism \citep{Cameron:1971lr}, and then resides to late times in a thick accretion disk around the equator of the RCB star. Li enhancement may only be detected if the star is viewed edge-on, directly probing this thick accretion disk. However, there is no observational evidence of disks in RCB stars, and the post-merger object will become spherical within a few dynamical timescales \citep{2012MNRAS.427..190S,2019MNRAS.488..438L,2019ApJ...885...27S}. \cite{2019MNRAS.488..438L} show that {\it MESA} models can exhibit Li on the surface of RCB stars without invoking a disk.

As shown in Figures~\ref{fig:all_abunds} and~\ref{fig:licnonef}, one of the biggest differences between the two different metallicities is the surface abundance of Li. For solar models, the Li abundance on the surface ($\log\epsilon(Li)$) is between 0.85 and 2.64, whereas the subsolar models have much higher Li abundances, between 3.95 and 6.41. The four observed Li abundances are between 2.6 and 3.5. The models that lie within the observations are marked in Table~\ref{tab:agree}. The Li abundance is also expected to be somewhat independent of
metallicity since it depends mainly on the $^3$He abundance in the progenitor WDs \citep{2012A&A...542A.117L}. Production of new Li depends on the Cameron-Fowler mechanism ($^3$He($\alpha$,$\gamma$)$^7$Be(e$^{-}$,$\nu$)$^7$Li) and the abundance of $^3$He. 

Our SOL models begin their post-merger evolution with a $^{3}$He mass fraction on the order of 10$^{-9}$ and a $^{7}$Li mass fraction on the order of 10$^{-9}$. These values agree with those from \cite{2014MNRAS.445..660Z} and \cite{2019MNRAS.488..438L}. However, our SUB models begin with a higher mass fraction of these elements, $^{3}$He mass fraction on the order of 10$^{-6}$ and $^{7}$Li mass fraction on the order of 10$^{-5}$. This difference is due to the way that {\it MESA} creates the He-WD. The test-suite function \texttt{make\_he\_wd} in {\it MESA} completes relaxation to WD phase by performing mass loss on the degenerate object until it reaches a set mass, which we set to 0.15 M$_{\sun}$. This mass is the same for both our SOL and SUB progenitors, and is scaled up to 0.25 M$_{\sun}$ when we create the RCB models. However, the He core in the SUB models is slightly smaller, and thus relaxing the He-WD to an equivalent mass as the SOL He-WD results in a small envelope that is slightly enriched in H, $^{3}$He, and $^{7}$Li. Thus, the progenitor for the SUB models is a He-WD with 0.024 M$_{\sun}$ hydrogen envelope that is slightly enriched in these elements.

While the difference between the SOL and SUB progenitors is very small, the effects are not negligible. Converting the quoted mass fractions to abundances, we find that the solar models begin with a surface abundance ($\log\epsilon(Li)$) of 2.3 and the subsolar models begin with $\log\epsilon(Li)$ of 6.3. Both of these values are very near to the upper bound of the RCB surface Li for our models. Curiously, our coldest models are at the lower bound of the range of surface Li, undergoing significant destruction of Li. The hotter models, then, must either have less Li consumption, or a similar amount of consumption accompanied with enhanced Li production due to the Cameron-Fowler mechanism. Additionally, we note that our reaction network, \texttt{mesa\_75.net} does not include the important reaction $^7$Li($\alpha$,$\gamma$)$^{11}$B. This reaction is 8 orders of magnitude more effective than the $\alpha$-capture on $^{14}$N \citep{Clayton2007}, which is paramount to the existence of surface $^{18}$O (see Sections~\ref{sec:CNO}--~\ref{sec:o18}). Munson et al. (in preparation) shows that the inclusion of $^{11}$B in the reaction network does significantly reduce the surface abundance of Li. Nevertheless, the RCB surface Li is strongly dependent on how much of the Li from the He-WD progenitor survives the WD merger, as it is certain that some Li would be destroyed in a merger event, the effects of which our models do not trace. The existence of surface Li in RCBs merits its own study.


\begin{table*}
	\centering
	\caption{Agreement with observations for each model and element. The rightmost column indicates how many criteria are met for each model.}
	\label{tab:agree}
	\begin{tabular}{lccccccccccr} 
		\hline
		Model ID & Li & C & $^{13}$C & N & O & $^{16}$O/$^{18}$O & C/O & F & Ne & Fe & total\\
		\hline
		SOL8.33 & no & no & no & yes & yes & no & no & no & yes & no & 3\\
		SOL8.39 & no & no & no & yes & yes & yes & no & no & yes & no & 4\\
		SOL8.44 & no & no & yes & yes & yes & yes & no & no & yes & no & 5\\
		SOL8.48 & yes & yes & yes & yes & yes & yes & no & no & yes & no & 7\\
		SOL8.53 & yes & yes & yes & no & yes & yes & no & yes & yes & no & 7\\
		SOL8.57 & yes & yes & yes & no & yes & yes & yes & yes & yes & no & 8\\
		SOL8.65 & yes & no & yes & no & yes & yes & no & yes & yes & no & 6\\
		SOL8.69 & no & no & yes & no & yes & no & no & yes & yes & no & 4\\
		SOL8.73 & no & no & yes & no & yes & no & no & yes & yes & no & 4\\
		\hline

		SUB8.33 & no & no & no & yes & no & no & yes & no & no & yes & 3\\
		SUB8.39 & no & no & no & yes & no & no & yes & no & no & yes & 3\\
		SUB8.44 & no & yes & yes & yes & yes & yes & no & no & no & yes & 6\\
		SUB8.48$^*$ & no & yes & yes & no & yes & yes & yes & yes & yes & yes & 8\\
		SUB8.53 & no & yes & yes & no & yes & yes & yes & yes & yes & yes & 8\\
		SUB8.57 & no & no & yes & no & yes & no & no & yes & yes & yes & 5\\
		SUB8.65 & no & no & yes & no & yes & no & no & yes & yes & yes & 5\\
		SUB8.69 & no & no & yes & no & yes & no & no & yes & yes & yes & 5\\
		SUB8.73 & no & no & yes & no & yes & no & no & yes & yes & yes & 5\\
		\hline
	\end{tabular}  
	
	$^*$ Denotes the preferred model
\end{table*}


\section{Discussion}
\label{sec:discussion}

In order to understand how RCB stars form and evolve, we must first understand the initial conditions created by the WD-merger events. In particular, we investigate two important parameters: the metallicity of the envelope, and the initial temperature of the He-burning shell, T\textsubscript{He}. The composition of the post-merger object depends on the metallicity of the progenitors. What sets the initial temperature of the He-burning shell is less well understood.

Previous studies have been guided by SPH and grid-based 3D hydro merger simulations. A number of q=0.7 simulations, which mimic an RCB-type WD merger, produced a range of ``Shell of Fire" (SOF) temperatures (analogous to our T\textsubscript{He}) of 1--2 $\times 10^8 K$ for both grid and SPH codes with and without AMR \citep{2018ApJ...862...74S}. The T\textsubscript{He} used in previous {\it MESA} models of RCB stars were based on these grid-based hydro simulations \citep{2013ApJ...772...59M,Menon_2018,2019MNRAS.488..438L}, and SPH simulations \citep{2011ApJ...737L..34L,2014MNRAS.445..660Z}. The T\textsubscript{He} assumed in these studies are summarized in Table~\ref{tab:t_he}. The values of T\textsubscript{He} in most of these studies are lower than those used for our models, but most do not include energy generation from nucleosynthesis and thus should be considered lower limits. The resulting T\textsubscript{He} from merger events depends on q and total mass such that a higher mass ratio merger produces a slightly lower SOF temperature, and a larger total mass system will produce a higher SOF temperature \citep{2018ApJ...862...74S}. 

There is evidence from the surface abundance of elements, such as Fe which is not affected by nucleosysnthesis, that the progenitor stars were metal poor. Similarly, the distribution of the RCB stars on the sky seems to indicate a bulge or old-disk population \citep{2012JAVSO..40..539C,2020A&A...635A..14T}.
The measured Fe abundances range from 5.5 to 6.8 while the solar value is 7.5. Thus, if the Fe abundance is an indication of metallicity then a reasonable value would be $\sim$1/10 Solar. Inspired by this, we computed a set of models with Solar (SOL) and Subsolar (SUB) progenitor metallicities to see how this variation affects the final surface abundances. 
It should be noted that there may be problems with assuming Fe is a metallicity indicator \citep{Lambert:1994uq,Asplund:2000qy}.

In the subsolar models, we see diminished abundances of all elements with the exception of C and Li. The C abundances are relatively unaffected by metallicity, as the extremely large abundance of He allows for constant replenishment of C due to the triple-$\alpha$ reaction, and the Li abundances, as we discussed in Section~\ref{sec:lithium}, are strongly dependent on the assumptions made about the progenitor He-WD.  
The diminished abundances (approximately one dex) of the other elements is expected as these models begin their evolution with 10$\%$ solar metallicity.
The abundances of N, O, and Ne roughly scale with metallicity. The $^{14}$N abundance is set by the progenitor metallicity, and, as discussed above, $^{18}$O and $^{22}$Ne are mostly formed from $\alpha$-captures on $^{14}$N \citep{2011MNRAS.414.3599J}.

 The consistency in the C abundances across the solar and subsolar metallicities leads to a strong difference in the C/O ratios. As shown in Figure~\ref{fig:CO_temp}, the solar models at cooler T\textsubscript{He} have C/O ratios that are far too small, with SOL8.44 and SOL8.48 having C/O = 0.12 and 0.09, respectively. However, at subsolar metallicities the O abundance has been diminished by approximately one dex while keeping the C abundances roughly the same. Thus, the C/O ratios are $\sim$1 for the cooler subsolar models, near the expected the ratio for RCB stars.

The choice of the initial Helium-burning shell temperature  (T\textsubscript{He}) is very important in determining the final surface abundances in our RCB star models.  The CNO abundances depend on correctly balancing the strength of the CNO cycle and He-burning at the base of the envelope. The most difficult of these three elements to replicate is the N abundance, which drops off steeply with increasing T\textsubscript{He}, as seen in Figures~\ref{fig:licnonef} and \ref{fig:N-O18-Ne-sum}.

We find that the temperature at which we get the best agreement for CNO is in the range of log(T\textsubscript{He}) $\sim$ 8.43 - 8.50. Our models SOL8.48 and SUB8.44 both have all three CNO elements lying within the observed values. The C isotopic ratio, $^{12}$C/$^{13}$C, is observed to be very large in most stars, and we are able to replicate this at log(T\textsubscript{He}) > 8.44, since the reaction $^{13}$C($\alpha$,n)$^{16}$O is the first $\alpha$-capture to occur at the onset of He-burning. This reaction is the dominant source of neutrons in the star, allowing for the formation of s-process elements, which are known to be enhanced in RCBs. Note that our reaction network does not include the formation of such elements. The observed $^{16}$O/$^{18}$O ratio is near unity, and our models replicate this behavior at intermediate temperatures, log(T\textsubscript{He}) $\sim$ 8.43 - 8.55. The C/O ratio, which defines carbon stars and governs the composition of dust grains, is observed to be greater than, but very nearly one. Our cooler models with log(T\textsubscript{He}) < 8.55 have small C/O ratios, however some of these models have C/O nearly zero, which is also not desirable. $^{19}$F is extremely overabundant in RCBs, and we find that this overabundance is reproduced at the highest temperatures, those with log(T\textsubscript{He}) > 8.5, where we have enough $\alpha$-captures going on to create $^{19}$F from $^{15}$N and $^{18}$O. The production of $^{18}$O is also strongly dependent on $\alpha$-captures. Ne is also overabundant in many of our models as it is in observations. This overabundance is reliably seen in models with log(T\textsubscript{He}) $\geq$ 8.48, as Ne is also an $\alpha$ element. 

Considering all information in Table~\ref{tab:agree}, the model which agrees most closely with the observations is SUB8.48. The only parameters which do not overlap with observations for this model are the N abundance, log$\epsilon$(N) = 6.67, which is very close to the observed range at only 0.5 dex lower than the minimum observation, and Li, which we mentioned could be adjusted by assuming a different progenitor abundance. We note that the model at one temperature step lower, SUB8.44, does put N in the observed area, however it doesn't reproduce the correct $^{19}$F or $^{22}$Ne abundances, and has a C/O ratio slightly below 1. The progenitor WDs will have gone through at least one common envelope phase in their evolution, and the effects of these common envelopes on the nucleosynthesis and abundances of close-binary systems is not well constrained. Therefore, while our preferred model is not perfect, it does match the observations remarkably well for the assumptions that have been made.

\begin{figure}
    \centering
    \includegraphics[width=\columnwidth]{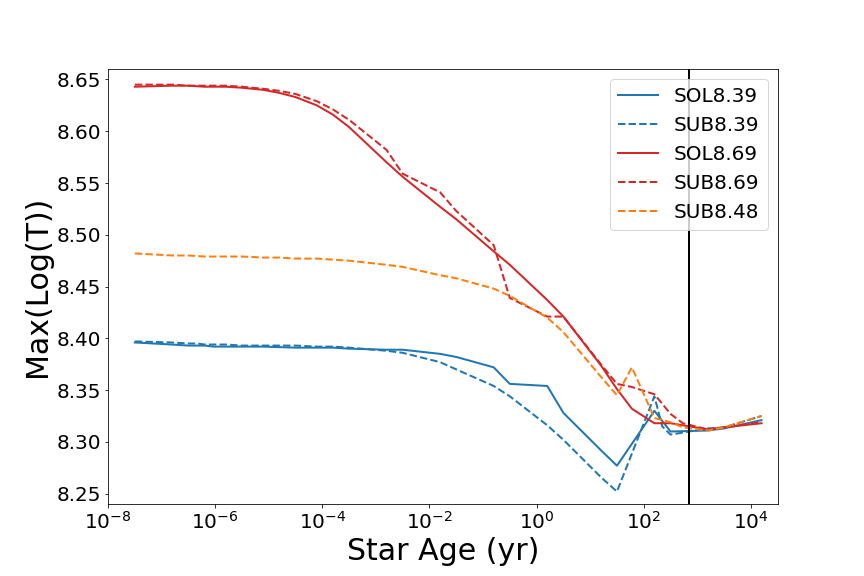}
    \caption{The trend of the maximum value of log(T) (the peak burning region temperature) within the star as a function of star age for four representative models and the preferred model. The solid lines indicate solar metallicities and the dashed lines indicate subsolar metallicities. The vertical black line indicates the average age for our models to reach RCB phase.}
    \label{fig:maxlogt}
\end{figure}
\begin{figure}
	\includegraphics[width=\columnwidth]{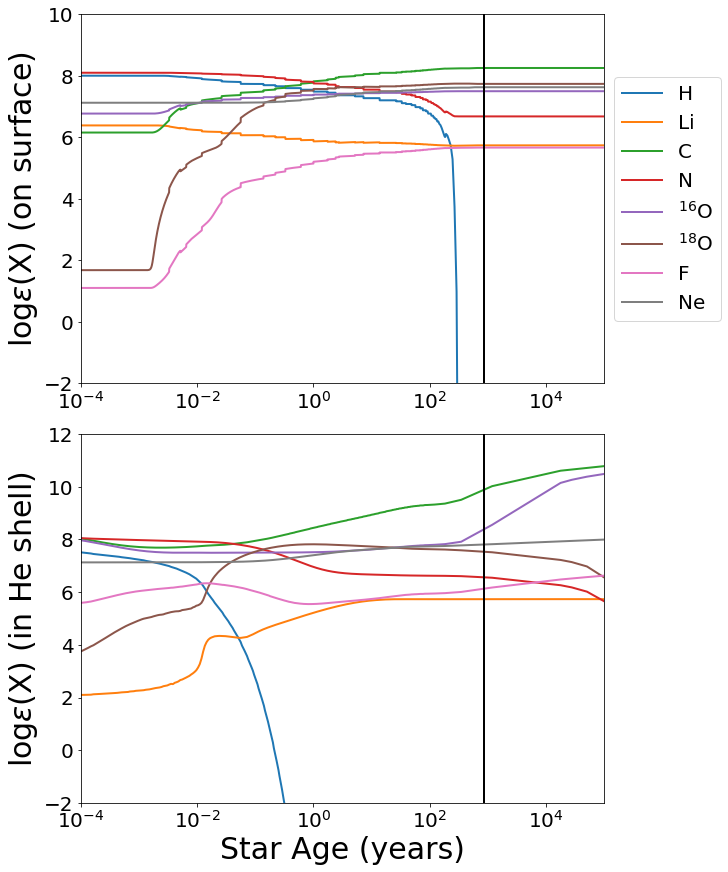}
    \caption{The evolution of abundances for the surface (upper panel) and the He-burning shell (lower panel) as a function of star age in years for our preferred model, SUB8.48. Each colored line represents a different element or isotope, and the vertical black line is the age where the RCB phase begins.}
    \label{fig:surface_changes_sol245}
\end{figure}
\begin{figure}
	\includegraphics[width=\columnwidth]{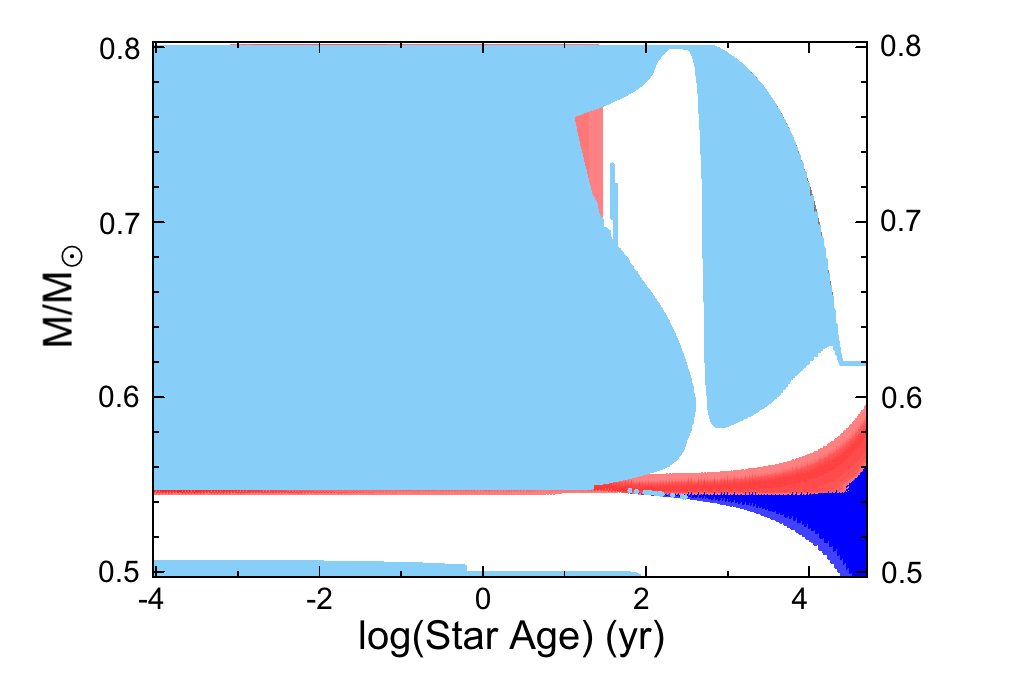}
    \caption{Kippenhahn diagram \protect\citep{2012sse..book.....K} for the preferred model, SUB8.48. The vertical axis is the mass coordinate, and the horizontal axis is the log of the star's age. Blue regions indicate convection at that mass coordinate and age and red regions indicate energy generation due to nucleosynthesis. The darker blue region indicates WD cooling in the core of the star. We note that there is nucleosynthesis in the outer envelope at early times, however we find its effect is minimal, as it generates on the order of 1 erg/g/s of energy.}
    \label{fig:kipp}
\end{figure}

The simplest explanation for the enrichment of RCBs in $^{18}$O and $^{19}$F is if these elements are the result of partial He-burning. There are two ways in which partial He-burning can occur. Either the He-burning shell only stays hot enough for a short period of time, or the partially-burned material is mixed out of the He-burning shell before it can be fully converted to its end products \cite{Clayton2007}. Our models exhibit both of these behaviors. As seen in Figure~\ref{fig:maxlogt}, the He-burning shell stays hot only for a few years, before cooling to an equilibrium temperature for all models at log(T\textsubscript{He}) = 8.31. It is, however, hot for a long enough timescale to produce large amounts of $^{18}$O and $^{19}$F. The upper panel of Figure~\ref{fig:surface_changes_sol245} shows for SUB8.48 that the surface abundances of $^{18}$O and $^{19}$F start increasing about 1 yr after the merger event and reach their equilibrium values after $\sim$10$^2$ yr, whereas the abundance of $^{18}$O in the He-burning shell peaks slightly before 1 yr and then begins to decrease, especially after the RCB phase is reached. The hot temperature of the He-burning shell creates these elements on a very short timescale. Figure~\ref{fig:kipp} shows the evolution of the convection zones with time in the evolving RCB star again for the SUB8.48 model. We see that the envelope is fully convective from the He-burning shell to the surface, beginning shortly after the merger event and lasting until $\sim$10 yr, at which point the inner and outer envelope split into two convective regions. After $\sim$10$^2$ yr, the convective zone pulls away from the He-burning shell and its material can no longer be mixed to the surface. Therefore the partial He-burning products, $^{18}$O and $^{19}$F, are created and mixed up to the surface within a short period of time. While the nucleosynthesis in the He-burning shell continues throughout the evolution as can be seen in the lower panel of Figure~\ref{fig:surface_changes_sol245}, there is no mechanism for its products to be lifted to the surface at late times, and thus the surface composition is constant. This interesting convective profile is calculated in {\it MESA} by the traditional mixing length theory \citep{1968pss..book.....C}, and while we do impose rotation on the models, we do not take rotational mixing into account. We ran a test model with rotational mixing turned on, and confirmed a result from \cite{2019MNRAS.488..438L}, that the addition of rotational mixing has a minimal effect on the surface abundances of the RCB model. In fact, our test model with rotational mixing included had identical RCB abundances to the model without it. The temperature profile in our post-merger objects is such that all convection happens organically through the evolution of the star, and there is no need for us to add in a mixing prescription separately, as was done in \cite{2013ApJ...772...59M}. Thus, these unique surface abundances of partial helium burning products are caused by the combination of a hot helium burning shell that quickly cools, and mixing that occurs after $^{18}$O and $^{19}$F are formed, but before they have time to be destroyed.

We find that our models take $\sim$10$^2$ years to reach the RCB phase, where they spend $\sim$10$^4$ years as an RCB, before evolving through the EHe phase in $\sim$10$^3$ years. The subsolar models evolve slightly slower than their solar counterparts.
The locus of the EHe stars is assumed to from the left (hot) side of the RCB locus to $\sim$40,000 K, or log(T\textsubscript{eff}) = 4.6. 
We calculate the lifetime of the EHe phase in our models as the difference between the age at log(T\textsubscript{eff}) = 4.6 and the age when the model leaves the RCB locus. This region of the HR diagram also includes some of the hottest known RCB stars, which only differ from EHe stars in that they exhibit declines in their light curves due to dust formation. 
The lifetimes are summarized in Table~\ref{tab:models}.  

 Using the birth rate from \cite{Karakas_2015} of 0.0018 yr$^{-1}$ combined with the model lifetimes, we calculate there would be around 30 RCBs and around 15 EHe stars in the galaxy. However, the current number of known RCBs in the Galaxy is 117 \citep{2020A&A...635A..14T} and we know of 22 EHe stars \citep{1996ASPC...96..471J,Jeffery_2017}. Assuming the birth rate is as quoted, we would need a longer RCB lifetime to match the population size that is observed. 
 
  We do, however, have a very good constraint of the real-time evolution of a hot RCB star, DY Cen. Archival plates have allowed us to watch the evolution of DY Cen from a cool RCB-like star in 1970 to a hot EHe-like star in 2014 \citep{2002AJ....123.3387D,2016MNRAS.460.1233S,2020MNRAS.493.3565J}. DY Cen has evolved through the EHe portion of the HR diagram, from roughly log(T\textsubscript{eff}) = 4.28 in 1987 to log(T\textsubscript{eff}) = 4.39 in 2015, in a timescale of about 30 years. Our models evolve through the same region of temperature space over an average timescale of 1750 years, significantly longer than DY Cen. 
  However, the contraction rates that have been estimated for EHe stars find that the mass plays a large role in the evolutionary speed of these stars \citep{1988MNRAS.235..203S,2002MNRAS.333..121S}. Since our models have rather low EHe masses (see Table~\ref{tab:models}) the contraction rates estimated by prior works point towards a much slower EHe evolution. Therefore, adjusting the mass loss in the RCB phase to lower values may in fact increase the EHe timescale of our models to something more closely resembling the evolution speed of DY Cen.
  As discussed in Section~\ref{sec:mesa}, the wind efficiency of these types of stars is not well constrained, and their effects on RCB and EHe lifetimes are complex. \cite{2019ApJ...885...27S} contains a nice discussion on the effects of mass loss on both the RCB lifetime and the ratio of RCB to EHe lifetime.

\section{Conclusions}
\label{sec:conclusions}

This is the latest in a series of studies using a combination of 3D hydro and 1D {\it MESA} simulations which have made significant progress in understanding how RCB stars form and evolve \citep{staff2012,2018ApJ...862...74S,2013ApJ...772...59M,2019MNRAS.488..438L}. 

By modulating the metallicity and initial He-burning shell temperatures of these RCB models, we are able to study the effects of these two important parameters. Remarkably, we are able to identify a single preferred model, SUB8.48, which has abundances closest to those of observed RCBs. This model is at 10\% of solar metallicity, and has an initial He-burning shell temperature of approximately 3.00 $\times 10^8 K$. We show that the convection of these models is such that the material exposed to the He-burning shell is mixed out of the He-burning region within the first few years after the merger event and brought to the surface where it can be observed. This gives one explanation as to why the RCB stars exhibit partial He-burning products on their surface. We're also able to explore the effects of T\textsubscript{He} and metallicity on the structure and evolution of RCBs. In general, subsolar metallicity RCBs have a higher surface temperature and thus a smaller radius, and live longer lives as RCBs. The difference in T\textsubscript{eff} is likely due to differences in opacity. The subsolar metallicity models experience a lower opacity, and are thus able to radiate more energy through the photosphere rather than having to spend its energy to expand the star. This effect can be seen as the subsolar models have smaller radii (and thus higher T\textsubscript{eff}), and slightly higher luminosities than the solar models. These two effects combine in the Bl{\"o}cker wind prescription \citep{1995A&A...297..727B} to decrease the mass loss, and thus extend their lifetime as RCBs.

We note that there are limitations on our estimates of RCB lifetimes, and thus population sizes, as these two depend strongly on the mass loss, which is not well constrained. Nevertheless, we calculate an average RCB lifetime on the order of 10$^4$ years and a population size of about 30 using a Bl{\"o}cker wind efficiency $\eta$ = 0.075, whereas the current number of known RCBs is nearly 120 in the Galaxy. Decreasing the wind efficiency of our models to $\eta$ = 0.005 increases the RCB lifetime by an order of magnitude, and increases the population size to around 250 RCBs, without changing the convective structure or the surface abundances.

There are still a few effects which we cannot explain well, or would need further exploration. While our models exhibit measurable Li on the surface of RCBs, we have not been able to replicate the observed abundances of this element, and there is reason to believe that the addition of $^{11}$B could destroy our remaining surface Li. However, our progenitor He-WD stars have an existing abundance of lithium, which plays a role in the amount seen on the surface during RCB phase. Future work is needed to make better assumptions on the lithium abundance of He-WD progenitors. 
We do not currently explore the effects of opacity in the models, but recent works such as \cite{2019ApJ...885...27S} have begun to explore that parameter space. Lastly, we acknowledge that {\it MESA} has limitations in regards to calculating the effects of a 3D merger process. We are now exploring whether a {\it MESA} model created by spherically averaging the 3D output of a hydrodynamical WD merger simulation is able to reproduce the results from stellar engineering models (Munson et al. 2020, in preparation).


\section*{Acknowlegements}
This work was supported by National Science Foundation Award 1814967. We would like to thank Amber Lauer, Josiah Schwab, Falk Herwig, Sagiv Shiber, and David Lambert for useful discussions. We would also like to thank our anonymous reviewer for insightful comments that helped us to strengthen this work. E.C. would like to thank the National Science Foundation for its support through award number AST-1907617 and the Louisiana State University College of Science and the Department of Physics \& Astronomy for their support.

\section*{Data Availability}
The MESA models generated in this study are available on request from the corresponding author.

\bibliographystyle{mnras}
\bibliography{everything2}

\bsp	
\label{lastpage}
\end{document}